\documentclass{epl}

\usepackage{graphicx}
\usepackage{dcolumn}
\usepackage{bm}

\title{The S$_0$(0) structure in highly compressed hydrogen and the orientational transition}
\shorttitle{The S$_0$(0) triplet in pressurized hydrogen}

\author{
Francesco Grazzi\inst{1}\thanks{E-mail: \email{grazzi@ifac.cnr.it}}
\and Massimo Moraldi\inst{2,3}\thanks{E-mail: \email{moraldi@fi.infn.it}}
\and Lorenzo Ulivi\inst{1,3}\thanks{E-mail: \email{ulivi@ifac.cnr.it}}
}
\shortauthor{F. Grazzi \etal}
\institute{
\inst{1} ISC--CNR, via della Madonna del piano, I-50019 Sesto Fiorentino, Italy \\
\inst{2} Dipartimento di Fisica, Universit\`a di Firenze, via G. Sansone~1, I-50019 Sesto Fiorentino, Italy\\
\inst{3} INFM, Unit\`a di Firenze, Via Giovanni Sansone~1, I-50019 Sesto Fiorentino, Italy
}

\begin{document}
\maketitle

\date{\today}

\begin{abstract}
A calculation of the rotational S$_0$(0) frequencies in high pressure solid para-hydrogen is performed.
Convergence of the perturbative series at high density is demonstrated by the calculation of second and third order terms.
The results of the theory are compared with the available experimental data to derive the density behaviour of structural parameters.
In particular, a strong increase of the value of the lattice constant ratio $c/a$ and of the internuclear distance is determined.
Also a decrease of the anisotropic intermolecular potential is observed which is attributed to charge transfer effects.
The structural parameters determined at the phase transition may be used to calculate quantum properties of the rotationally ordered phase.
\end{abstract}


The phase diagrams of hydrogen and deuterium demonstrate amazing phenomena caused by the quantum nature of both translations and rotations of the molecules in these solids \cite{Silvera80,Mao94}.
At low pressure, both systems have an hexagonal close packed (hcp) structure, with almost freely rotating molecules \cite{VanKranendonk83}.
A transition to a broken symmetry phase where the molecules are oriented, and the equivalence of the sites is lost (BSP or phase II) occurs for low temperature at 28 GPa for D$_2$ and 110 GPa for H$_2$. \cite{Silvera81,WijngaardenPhD82,Lorenzana90}
An appropriate description of the BSP is still lacking.
The correct quantum treatment of nuclei is indeed crucial for the description of the BSP transition, given the extremely large isotopic effect on the transition pressure.
For this reason a description on the basis of a molecular interaction potential offers numerous advantages with respect to, for example, \textit{ab initio} molecular dynamics \cite{Kohanoff97} (which generally treats the nuclear motion as classical), quantum Monte Carlo techniques (limited to zero temperature), \cite{Ceperley87} or other more sophisticated computation techniques, still prohibitive for heavy computer time needed \cite{Kitamura00}.

An effective potential model for the anisotropic molecular interaction in the solid has never been derived, especially because a reliable theory to derive solid spectroscopic data was lacking.
In this paper we present a theory able to calculate the roton frequencies in solid hcp para-hydrogen up to very high pressure.
From the comparison with experimental data, we obtain information on the anisotropic components of the hydrogen-hydrogen intermolecular potential, and we derive, as a function of pressure, fundamental structural parameters as the lattice constant ratio $c/a$ and the internuclear distance.
Here we give a further demonstration that effective two-body potentials are well suited to explain experimental results in high pressure systems and, as such, they are possibly able to explain the transition to the rotationally ordered phase.

The triplet structure of the S$_0(0)$ Raman transition for para-hydrogen is known since long time \cite{Bhatnagar62}, and is considered a clear evidence of the rotationally disordered hcp lattice.
Its pressure evolution, including its interaction and hybridisation with the phonon, occurring at intermediate densities, has been measured up to the transition to the BSP \cite{Silvera81,WijngaardenPhD82,Lorenzana90}, and has been the subject of a renewed interest also recently \cite{Goncharov01}.

At low density a first-order perturbative theory is usually employed for the theoretical description of the triplet.
In such a picture the first excited rotational state ($J=2$) broadens in a band of collective states (rotons).
Raman transitions from the ground state are allowed only to five states which correspond to $\mathbf{q}=0$ rotons belonging to symmetry representations $A_{1g}$, $E_{1g}$ and $E_{2g}$ of the rotation group of the crystal.
The $E$ states are doubly degenerate and thus the Raman spectrum results in a triplet \cite{VanKranendonk83}.
The comparison of the first order perturbation theory with experiment has not been satisfactory and the applicability of perturbative methods to the compressed solid has been questioned \cite{Loubeyre91,Goncharov01}.
In the following we demonstrate that, going beyond first order, the perturbation expansion is convergent and the numerical results are reliable up to the density of the BSP transition.

We consider a rigid hcp lattice of rigid molecules.
The zeroth order eigenfunctions of the $J=2$ Raman active ($\mathbf{q}=0$) roton modes are given by 
\begin{eqnarray}
\label{eigen}
\Psi_{\mu} \sim  \sum_i^{2N} Y_{2\mu}(\omega_i)
\end{eqnarray}
where $i$ runs over the molecules and $N$ is the number of unit cells, $\mu$ can take the values 0, $\pm$1 and $\pm$2 that correspond to the A$_{1g}$, E$_{1g}$, E$_{2g}$ representations respectively.
The spherical harmonics $Y_{Lm}$ are calculated with respect to a reference system having the $z$-axis coincident with the $c$-axis of the hcp crystal.
The anisotropic intermolecular potential is assumed to be pair-additive, and is expanded in spherical components in the usual way (see eqs.~10 and 11 of ref.~\cite{Grazzi02}).
We consider here those spherical components that have been evaluated in the literature, \cite{Norman84,Schaefer89,Diep00} namely $V_{202}$ ($ = V_{022}$) that takes into account crystal field effects, and $V_{22L}$ (with $L=0, 2, 4$) that is responsible for rotonic states.

First order energies for the excited state are written \cite{VanKranendonk83} 
\begin{equation}
\epsilon_{\mu }^{(1)} = (-1)^{\mu }\sum_{L}C(22L;\mu ,-\mu ,0)  \left[S_{22L}+5(2L+1)^{-\frac{1}{2}} C(22L;0,0,0)S_{L0L}\right] 
\end{equation}
where $C(L_1\,L_2\,L;m_1,m_2,m)$ are the Clebsch-Gordan coefficients, and $S_{l_{1}l_{2}L}$ are the lattice sums 
\begin{equation}
S_{l_{1}l_{2}L}=\sqrt{4\pi }\sum_{j\neq 1}V_{l_{1}l_{2}L}(R_{1j})Y_{L0}(\Omega _{1j}).
\end{equation}
The first order correction removes the degeneracy among states with different $|\mu|$, giving rise to three Raman active transitions of frequencies $\nu_{|\mu|}$,with $\nu_0 > \nu_2 > \nu_1$.
The splitting is symmetric ($\nu_0 -\nu_2 = \nu_2- \nu_1$ ) only if calculated at first order and neglecting all but the $V_{224}$ potential components.

The general expression for second order corrections to the $J=2$ state is \cite{Landau74}
\begin{eqnarray}
\epsilon^{(2)}_{\mu}= \sum_s \frac{|\langle \Psi_{\mu}| V | \Phi(s) \rangle|^2}{E_s -E_2} = \sum_s \epsilon^{(2)}_{\mu} (s)
\end{eqnarray}
where $s$ indicate any rotational state other than the single $J=2$ excitation and $\Phi(s)$ and $E_s$ are the zeroth order wave-functions and energies for the rotational state $s$.
The second equality is a definition of $\epsilon^{(2)}_{\mu}(s)$.
We have also calculated the expressions for the third order corrections both to the ground state and to the $J=2$ rotational state.
The calculation involves now two intermediate rotational states $s_1, s_2$ \cite{Landau74} and the different terms are indicated with $\epsilon^{(3)}_{\mu}(s_1;s_2)$ accordingly.

Going beyond first order one finds that also the energy of the ground state changes, with a correction that is proportional to the number of molecules $2N$, having therefore an extensive character.
Anyway, both the second and third order corrections to the energy of the excited state contain quantities proportional to $2N$, that are identical to the corresponding corrections of the ground state and thus cancel out when calculating the transition frequencies $\nu_\mu$.
For the frequencies at second order $\nu^{(2)}_\mu$ the terms that contribute are 
\begin{eqnarray}
h\nu^{(2)}_\mu = \epsilon^{(2)\prime}_{\mu}(22)+\epsilon^{(2)\prime}_{\mu}(222)+\epsilon^{(2)}_{\mu}(42) + \epsilon^{(2)}_{\mu}(40) 
\label{socorr}
\end{eqnarray}
where the prime means that the extensive contributions have been subtracted and the argument of $\epsilon$ here explicitly indicates the number and nature of rotational excitations of the intermediate state $s$ ({\it e.g.} 222 indicates 3 rotational excitations with $J=2$).
The terms appearing in Eq.~\ref{socorr} depend on products of the lattice sums $S_{202}$ and $S_{22L}$ and on the new lattice sums $T_{LL^\prime M}$ \cite{Grazzi04}
\begin{equation}
T_{LL^\prime M} = 4\pi \sum_{j\ne 1} V_{22L}(R_{ij})   V_{22L^\prime}(R_{1j})  
Y_{LM}(\Omega_{1j}) Y^{\ast}_{L^\prime M}(\Omega_{1j})  .
\end{equation}

At third order the terms that contribute are
\begin{eqnarray}
h\nu_{\mu }^{(3)} & = &\epsilon_{\mu}^{(3)\prime}(22;22) +\epsilon _{\mu }^{(3)\prime }(222;222)  +\epsilon _{\mu }^{(3)\prime }(42;42) + \nonumber  \\
&+& \epsilon_{\mu }^{(1)} \left[\frac{\epsilon _{\mu }^{(2)\prime}(22)}{\Delta_{20}} +\frac{\epsilon _{\mu }^{(2)\prime }(222)}{2\Delta _{20}}+\frac{\epsilon_{\mu}^{(2)}(42)}{\Delta _{40}} \right]+ \nonumber \\
&+& \epsilon _{\mu }^{(3)\prime }(222;22)+\epsilon _{\mu }^{(3)\prime}(222;42)+\epsilon _{\mu }^{(3)\prime }(22;42)
\label{treord}
\end{eqnarray}
where $\Delta _{40}$ and $\Delta _{20}$ are zero order energy differences of rotational states.
The calculation requires the computation of double and triple lattice sums \cite{Grazzi04}.
For the third order correction only (Eq. \ref{treord}) we have neglected crystal field effects $V_{202}$.
This is justified by the smallness of third order effects (see below).

By the use of the analytical expressions for $V_{224}$ and $V_{202}$ given by \cite{Norman84} and the interpolation of the table given in \cite{Schaefer89} for $V_{220}$ and $V_{222}$ we have calculated the frequencies of the S$_0$(0) triplet, and analysed independently the first, second and third order correction terms.

Even if the first order term exceeds 100 cm$^{-1}$ at high density, the perturbative expansion is convergent up to 340 mol/l and more, since the third order correction is always less than 2 \% of the unperturbed value $\nu_{gas} = 354.4$ cm$^{-1}$.
Writing explicitly the corrections in a typical high density case ($\rho=300$ mol/l, $c/a=\sqrt{8/3}$), we have $\nu_0=\nu_{gas}+94 + 14 +5 $, $\nu_1=\nu_{gas}-36  +43  -2 $ and $\nu_2=+29 +34 +1 $ cm$^{-1}$.
We have found that, even though the main contribution arises from the $V_{224}$ potential component, which comprises the electric quadrupole-quadrupole (EQQ) interaction, the effect of the other potential terms are relevant ($\simeq 20$ cm$^{-1}$) if $c/a$ deviates a few percent from the ideal value.
The effect of the crystal field term, $V_{202}=V_{022}$ is negligible (only 0.02 \%) for $c/a=\sqrt{8/3}$.
At high density, varying $c/a$ of 3 \% may produce a change of 30 cm$^{-1}$ for the absolute value of the frequencies and of about 44 \% of the splitting.
The $c/a$ ratio is therefore an important parameter to be considered when trying to reproduce the experimental rotational frequencies.

\begin{figure}[thb]
\begin{center}
\includegraphics[bb= 2.0cm 6.0cm 20cm 25cm, width=10cm]{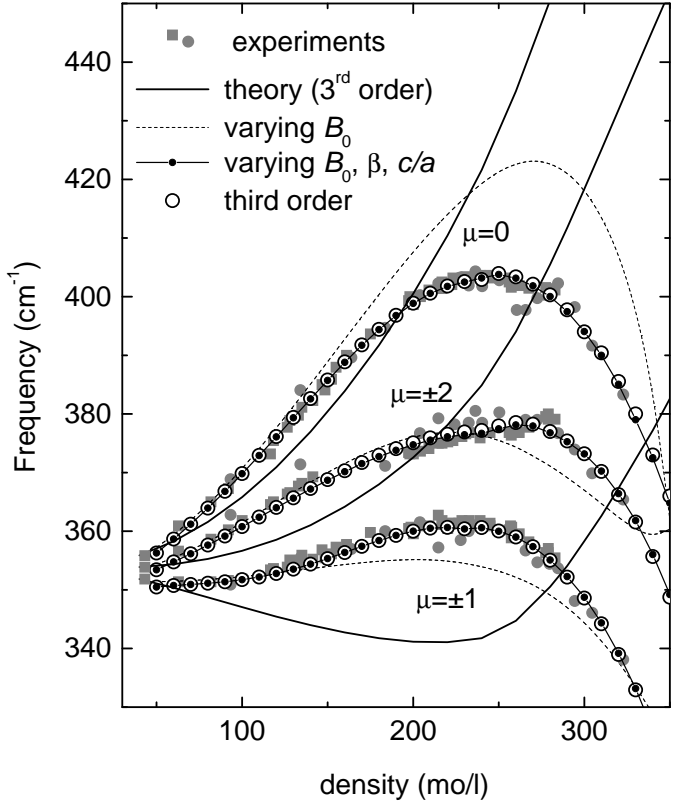}
\caption{Theoretical results with fixed parameters (solid lines) and comparison of the results of the calculation with the experiment.
Solid grey squares and dots are the experimental data for hydrogen of ref.s~\protect\cite{Silvera81,Goncharov01}.
The dashed line and solid dots joined by a line represent the values calculated with different free parameters (see text).
The empty circles are the values calculated including third order terms, with parameters corresponding to the solid dots. 
\label{f.Freq} }
\end{center}
\end{figure}

It is important to notice an increase of the calculated average energy of the triplet at high density (this is true even considering only first order term if $V_{220}$ is included in the calculation) and an asymmetric reduction of the splitting.
Results of the calculation, including the third order, with fixed rotational constant $B_0=59.246$ cm$^{-1}$ are reported in fig.~\ref{f.Freq}. 
We have compared our theoretical results with experimental data \cite{WijngaardenPhD82,Silvera81,Goncharov01,Lorenzana90}, also reported in fig.~\ref{f.Freq}, as a function of density $\rho$ calculated from the reported pressure by using the room temperature equation of state, neglecting thermal pressure effects \cite{Loubeyre96b}.
The experimentally observed overall decrease of the rotational frequencies above a certain density contrasts with the increase of the theoretical values and addresses for a decrease of $B_0$ with density.
It is also evident that above 150 mol/l the calculated splitting overestimates the experimental one.
This problem was already evident to other authors who have used only first order expansion and only EQQ interaction \cite{Loubeyre91,Goncharov01}.
Such a discrepancy in one case has been ascribed to the believed failure of the perturbative expansion above 40 GPa, and it is corrected below 40 GPa assuming a screening of the EQQ interaction in the solid, proportional to the square root of the refractive index \cite{Loubeyre91}.
By contrast, other authors assume that the first order perturbation theory is sufficient up to high pressure, and introduce a new crystal field term, that they calculate at each pressure ascribing to this one the whole discrepancy \cite{Goncharov01}.
Both approaches are not completely convincing, because they are limited by the first order perturbative theory.
We are in a much better position, since, due to the convergence of the perturbative series, we can confidently push our analysis to the whole experimentally investigated density range, considering the absolute value of the individual roton frequencies and not only the splitting.

%
Our conclusion is that, in addition to $B_0$, also the anisotropic interaction energy has to be assumed to change significantly with increasing density in the solid phase.
We have modelled this change in an effective way, by multiplying all the potential components by the same factor, $\beta$.
We have pushed our analysis even further, taking full advantage of the information contained in the experimental data of the triplet frequencies, to derive, at any density, the third unknown parameter entering in the calculation, namely, the lattice constant ratio $c/a$.
The results of the analysis are reported in fig.~\ref{f.Freq}.
Letting only $B_0$ to vary ($c/a=\sqrt{8/3}$, $\beta=1$ ) we cannot find an agreement for the individual frequencies, since the splitting is still overestimated (dashed line), while we can perfectly reproduce the data by adjusting $B_0$, $\beta$, and $c/a$.
The analysis has been performed using the expansion up to second order (solid line with dots), and checking consistency summing up the third order contribution (open circles).

%
\begin{figure}[htb]
\twofigures[bb= 3.0cm 3.0cm 20cm 26cm, width=7cm] {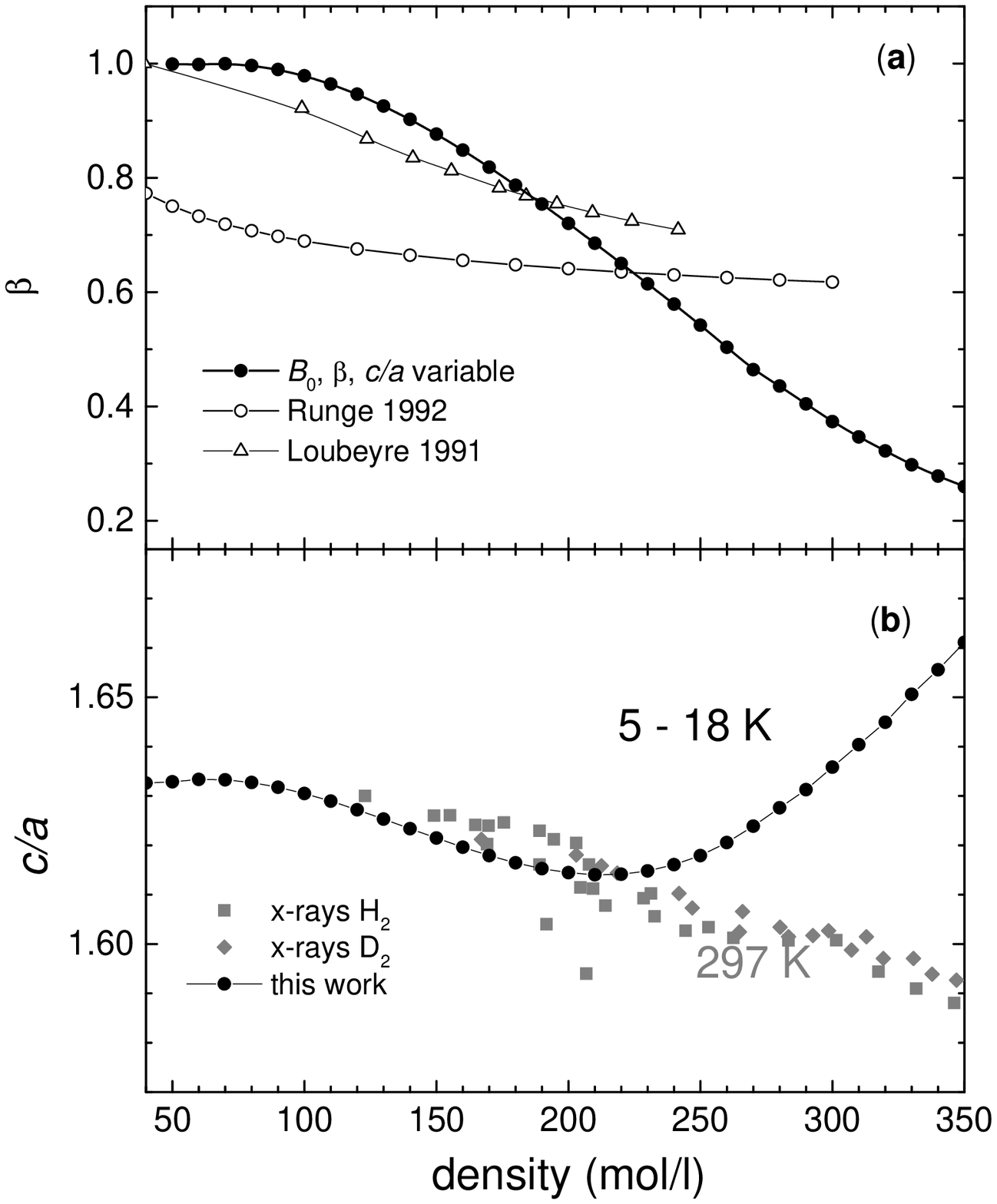}{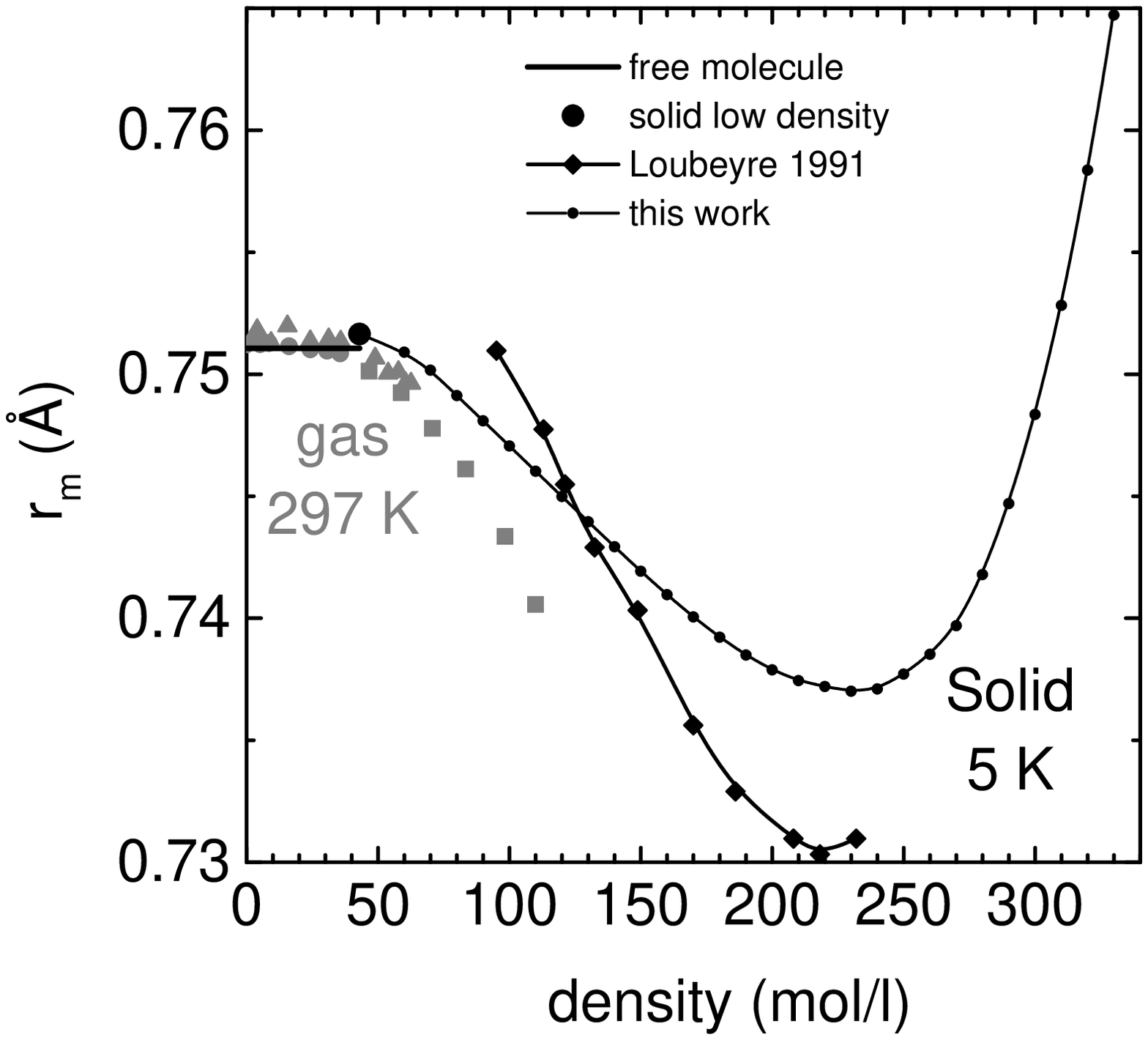}
\caption{({\bf a}) Attenuation factor $\beta$ versus density.
The values used in ref.~\protect{\cite{Loubeyre91}} (triangles) and in ref.~\protect{\cite{Runge92}} (circles) are reported.
 ({\bf b}) Lattice constants ratio $c/a$. 
The grey symbols are the room temperature x-rays determination \protect\cite{Loubeyre96b}.
\label{f.fattetc}}
\caption{Bond length for H$_2$ as a function of density.
The grey symbols are values determined in the gas at room temperature (dots ref.~\protect{\cite{May61}}, triangles ref.~\protect{\cite{Moulton88}} and squares ref.~\protect{\cite{Ulivi98}}); diamonds are data derived in ref.~\protect{\cite{Loubeyre91}}.
The big solid dot is the value derived from data in ref.~\protect{\cite{Bhatnagar62}} and the horizontal grey lines represent values derived from the rotational constants in the low density gas.
\label{f.Req} }
\end{figure}

The density behaviour of $\beta$ and $c/a$ are reported in fig.~\ref{f.fattetc}.
Even though an undoubtedly large effect ($\beta \simeq 0.3$ at high density), the attenuation of the pair anisotropic potential is not surprising.
Moreover, the 70 \% variation we have found is, in absolute value, much smaller than the difference between the gas \cite{Schaefer89} and the solid \cite{Silvera78} isotropic potential.
Effective anisotropic interaction has been occasionally used to calculate properties of high pressure hydrogen, and in particular the pressure where the transition to the BSP is expected.
Using the gas value of the EQQ interaction has always produced too low values for the transition pressure, indicating an overestimation of the anisotropic interaction \cite{England76,Janssen90}.
For HD, where the transition line has a particular reentrant behaviour \cite{Moshary93}, a correct estimation of the transition pressure can be obtained by a mean-field theory and considering the effect of a crystal field term that may be physically produced by a lattice distortion \cite{Freiman98c}. 
In fig.~\ref{f.fattetc}a we report also the attenuation factor assumed by Loubeyre \textit{et al.} \cite{Loubeyre91} and that derived by Runge \textit{et al.} \cite{Runge92} and used in a Path-Integral Monte Carlo calculation\cite{Cui97}.
Both of them are of the same order of magnitude as ours.

For what concerns the lattice constant ratio $c/a$, reported in fig.~\ref{f.fattetc}b, we notice that, up to $\rho\simeq$200 mol/l the spectroscopically determined value differs only slightly from either $\sqrt{8/3}$ or the room temperature x-rays value, while, above, a marked increase is evident.
A recent self-consistent analysis of lattice distortion concludes, on the basis of a mean-field approach, that the hcp lattice remains ideal up to the BSP transition pressure, where a large variation of $c/a$ occurs \cite{Freiman01,Freiman02}.
This increase occurs for different densities in the cases of H$_2$ and D$_2$ \cite{Grazzi04}, thus demonstrating that it represents a precursor of the phase transition, rather than a mere density effect.
At room temperature, indeed, this increase is not present \cite{Loubeyre96b}, while an increase of the lattice constant $c$ passing from the hcp to the BSP phase in D$_2$ at constant pressure has been noticed in a x-ray diffraction measurement \cite{LoubeyreVarenna}.

In fig.~\ref{f.Req} we report the effective internuclear distance $r_m=\left\langle r^{-2} \right\rangle ^{-1/2}= (4 \pi \mu c B_0/\hbar)^{-1/2} $.
In the same figure, we show data referring to the gas at room temperature \cite{May61,Moulton88,Ulivi98}, to the isolated molecule \cite{Foltz66} and to the low pressure solid \cite{Bhatnagar62}.
The marked decrease of the internuclear distance, already seen in ref.~\cite{Loubeyre91}, proceeds only up to about 220 mol/l, where a rapid increase takes place.
At the BSP transition, it becomes even larger than the isolated molecule value, probably favoured by the onset of the orientational ordering.
Our results show that the rise of $r_m$ and $c/a$  are two important features to be interpreted as precursors of the BSP transition.
The increase of $r_m$ (with an expected increase of the molecular quadrupole moment [3]) and the decrease of the anisotropic potential seem contradictory due to the eminently quadrupolar character of the interaction responsible for the S$_0(0)$ structure.
Such a contradiction calls for the consideration of charge transfer effects which are present even at low pressure and are especially important for determining the anisotropic intermolecular interaction.
The structural parameters and interaction potential at the transition should enable one to calculate equilibrium structure and dynamical effects in the BSP.
%

\end{document}